\documentclass{uncecomp2017}
\usepackage{amsmath}
\usepackage{bm}
\usepackage{dsfont}
\usepackage{amsfonts}
\usepackage{graphicx}
\usepackage{subcaption}
\usepackage{cleveref}
\usepackage{breqn}
\usepackage{algorithm,algorithmic}

\title{Multilevel preconditioning of Polynomial Chaos Method for quantifying uncertainties in a blood pump}

\author{Chen Song$^{1,2}$, Vincent Heuveline$^{1,2}$}

\heading{Chen Song, Vincent Heuveline}

\address{$^1$Heidelberg Institute for Theoretical Studies\\
  Schloss-Wolfsbrunnenweg 35, 69118 Heidelberg, Germany\\
  e-mail: \{chen.song, vincent.heuveline\}@h-its.org \and
  $^2$
  Engineering Mathematics and Computing Lab (EMCL)\\
  Interdisciplinary Center for Scientific Computing (IWR), Heidelberg Univeristy\\
  Im Neuenheimer Feld 205, 69120 Heidelberg, Germany\\
  e-mail: \{chen.song, vincent.heuveline\}@iwr.uni-heidelberg.de}


\abstract{
Heart failure (HF) is a severe cardiovascular disease, it happens when the heart muscle is so weakened such that it can not provide sufficient blood as body needs. More than 23 million people are suffered by HF worldwide. Despite the modern transplant operation is well established, the lack of heart donations becomes a big restriction on transplantation frequency. With respect to this matter, ventricular assist devices (VADs) can play an important role in supporting patients during waiting period and after the surgery.

Moreover, it has been shown that VADs by means of blood pump have advantages for working under
different conditions. While a lot of work has been done on modeling the functionality of the blood pump, but quantifying uncertainties in a numerical model is a challenging task. We consider the Polynomial Chaos (PC) method, which is introduced by Wiener for modeling stochastic process with Gaussian distribution. The Galerkin projection, the intrusive version of the generalized Polynomial Chaos (gPC), has been densely studied and applied for various problems. The intrusive Galerkin approach could represent stochastic process directly at once with Polynomial Chaos series expansions, it would therefore optimize the total computing effort comparing with classical non-intrusive methods. We compared different preconditioning techniques for a steady state simulation of a blood pump configuration in our previous work, the comparison shows that an inexact multilevel preconditioner has a promising performance. In this work, we show an instationary blood flow through a FDA blood pump configuration with Galerkin Projection method, which is implemented in our open source Finite Element library Hiflow3. Three uncertainty sources are considered : inflow boundary condition, rotor angular speed and dynamic viscosity, the numerical results are demonstrated with more than 30 Million degrees of freedom by using supercomputer.
}

\begin{document}

\section{INTRODUCTION}

Since last two decades, medical instruments have been playing an important role in the field of health care, they sustain patients' life by providing advanced clinical solutions or help surgeons to obtain better medical diagnosis. The innovation of medical devices expanded significantly recent years owing to the improvement of mechanical design, numerical modeling, computing hardware, etc. Especially, the blood pump device became one of the most popular assist apparatuses in heart failure treatment \cite{Birks2006a, Burkhoff2006a, Pagani2009a}. Ventricular assist devices (VADs) can be applied as long-term implant for patients who are not eligible for heart transplantation, or served as a generation of sufficient blood flow for patients during the surgery. However, the verification and validation (V\&V) procedure is still demanding due to the lack of knowledge about input parametric data, specially for numerical modeling. Whereas, these uncertainties should not be ignored, because quantifying the impact of these uncertainties can be one of the key points for patient-specific implantation. 

In our previous work \cite{Schick2015a}, we investigate the steady state of Navier-Stokes equations for laminar flow in a blood pump geometry. We modeled the rotating impact with the Multiple Reference Frame (MRF) method \cite{Bujalski2002a, Krause1968a, Cruz2003a}, for quantifying uncertainties, we employed the intrusive Polynomial Chaos Expansion (PCE) method in order to take geometrical uncertainty and specific model parameters into account. We placed also our attention on comparing different solving strategies for the intrusive Polynomial Chaos approach for fluid problem, as being capable of solving efficiently this highly coupled system is very crucial in practice.

In this paper, we will analyze an unsteady blood flow in a blood pump device based on the Variational Multi-Scale method (VMS) for the incompressible Navier-Stokes equations. We concentrate again three different uncertainty sources: inflow boundary condition, dynamic viscosity and angular speed of the rotor. The geometry of the blood pump is referred to the U.S. Food and Drug Administration (FDA) "critical path" initiative benchmark problem \cite{FDA}, the aim of this project is to utilize advanced computational simulations to predict the biological response. So as to cope with the rotating machinery modelization, we employ a shear layer update approach, which is very similar to The Shear-Slip Mesh Update Method (SSMUM) \cite{Behr1999a, Behr2003a}, the main difference is to deploy two layers instead of one in order to facilitate the update process once a regeneration of the mesh is needed. By this setting, the local linear algebra structure remains the same, hence, only few values demand update. We employ the Variational Multi-Scale method \cite{Dubois2004a, McDonough1999a}, which inherits the consideration of two separated scales from the large eddy simulation (LES) model and the concept of stabilized finite element method \cite{Tezduyar1991a}, it provides the feasibility of modeling the blood flow within the high rotation speed instrument. 

We here apply the stochastic finite element method (SFEM) \cite{Ghanem1991a}, this conveys that the spatial domain is discretized by the finite element method (FEM), and the stochastic space is expressed by Polynomial Chaos Expansion(PCE) \cite{Wiener1938a}. One important advantage of utilizing PCE is when the model output is smooth regarding to the input, a spectral convergence of PCE is achieved \cite{Crestaux2009a}. PCE exhibits the stochastic solution via the orthogonal multivariate polynomials, and the input random variables are represented in predefined polynomials regarding to their probability distribution. The stochastic Galerkin projection \cite{Ghanem1991a, LeMaitre2010a} is a powerful tool, which provides the possibility to obtain the coefficients of the PCE system at once. However, efficient solving strategies are needed for such complex structure. From our previous contribution \cite{Schick2015a}, we compared three different preconditioning techniques for a stationary flow in a simplified blood pump geometry, the inexact Multilevel method outperformed the Mean based preconditioner and the exact Multilevel method. We focus therefore in this paper on the inexact Multilevel method, and apply it to the instationary flow in the pump geometry.

The rest of this paper is organized as follow. In \Cref{sec:math}, we introduce the mathematical modeling for high Reynolds number flow in a rotating machinery, illustrate our moving mesh technique and the modelization of stochastic system. \Cref{sec:num_method} presents our Multilevel preconditioner algorithm. The numerical results are showed in \Cref{sec:num_result}. We conclude our contribution and provide an outlook for further development in \Cref{sec:conclusion}.

\section{MATHEMATICAL MODELING}
\label{sec:math}

We consider the unsteady state of the incompressible Navier-Stokes equations in a rotating machinery, it yields:

\begin{subequations}
\begin{align}
\frac{\partial \bm{u}}{\partial t} + ((\bm{u} - \bm{u}^r) \cdot \nabla) \bm{u} - \frac{\mu}{\rho} \Delta \bm{u} + \frac{1}{\rho} \nabla p &= 0 \;, &&\text{in} \: \Omega \;,\\
\nabla \cdot \bm{u} &= 0 \;, &&\text{in} \: \Omega \;,\\
\bm{u}^r &= \bm{d} \times \bm{\omega} \;, &&\text{in} \: \Omega_{rot} \;,\\
\bm{u}^r &= 0 \;, &&\text{in} \: \Omega_{stat} \;,\\
\bm{u} &= \bm{g} \;, &&\text{on} \: \Gamma_{in} \;,\\
(\frac{\mu}{\rho} \nabla \bm{u} -p\mathds{1}) &= 0 \;, &&\text{on} \: \Gamma_{out} \;,\\
\bm{u}^r &= \bm{d} \times \bm{\omega} \;, &&\text{on} \: \Gamma_{rotor} \;,\\
\bm{u} &= 0 \;, &&\text{on} \: \partial \Omega \backslash (\Gamma_{in} \cup \Gamma_{out} \cup \Gamma_{rotor}) \;.
\end{align}
\label{eq:ns}
\end{subequations}

\noindent Where, $\bm{u}$ is the velocity, and $p$ is the pressure. The flow is described on $\Omega \subset \mathbb{R}^3$, $\Omega = \Omega_{rot} \cup \Omega_{stat}$, $\Omega_{stat} \cap \Omega_{rot} = \emptyset$. $\Omega_{stat}$ and $\Omega_{rot}$ denote static and rotating domain respectively \cite{Behr2003a}. $\bm{u}^r$ is the rotation speed, which pre-describes the motion of a moving mesh, it is defined by the angular speed $\bm{\omega}$ and the distance to the rotating axis $\bm{d}$. $\mu$ is the dynamic viscosity and $\rho$ is the density. We ignore here the external body force $\bm{f}$ in N-S equations.

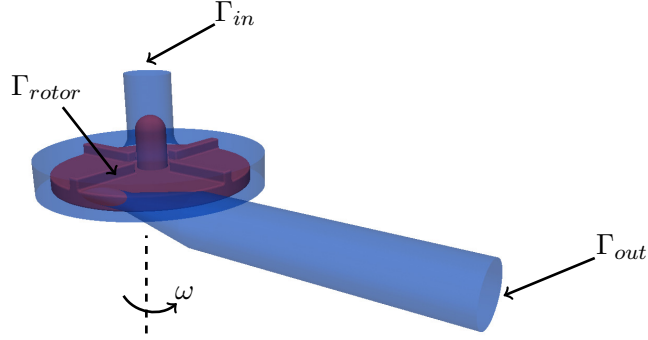
\begin{figure}
	\centering
	\begin{tikzpicture}
	\node[anchor=south west,inner sep=0] (image) at (0,0)
	{\includegraphics[width=0.5\textwidth]{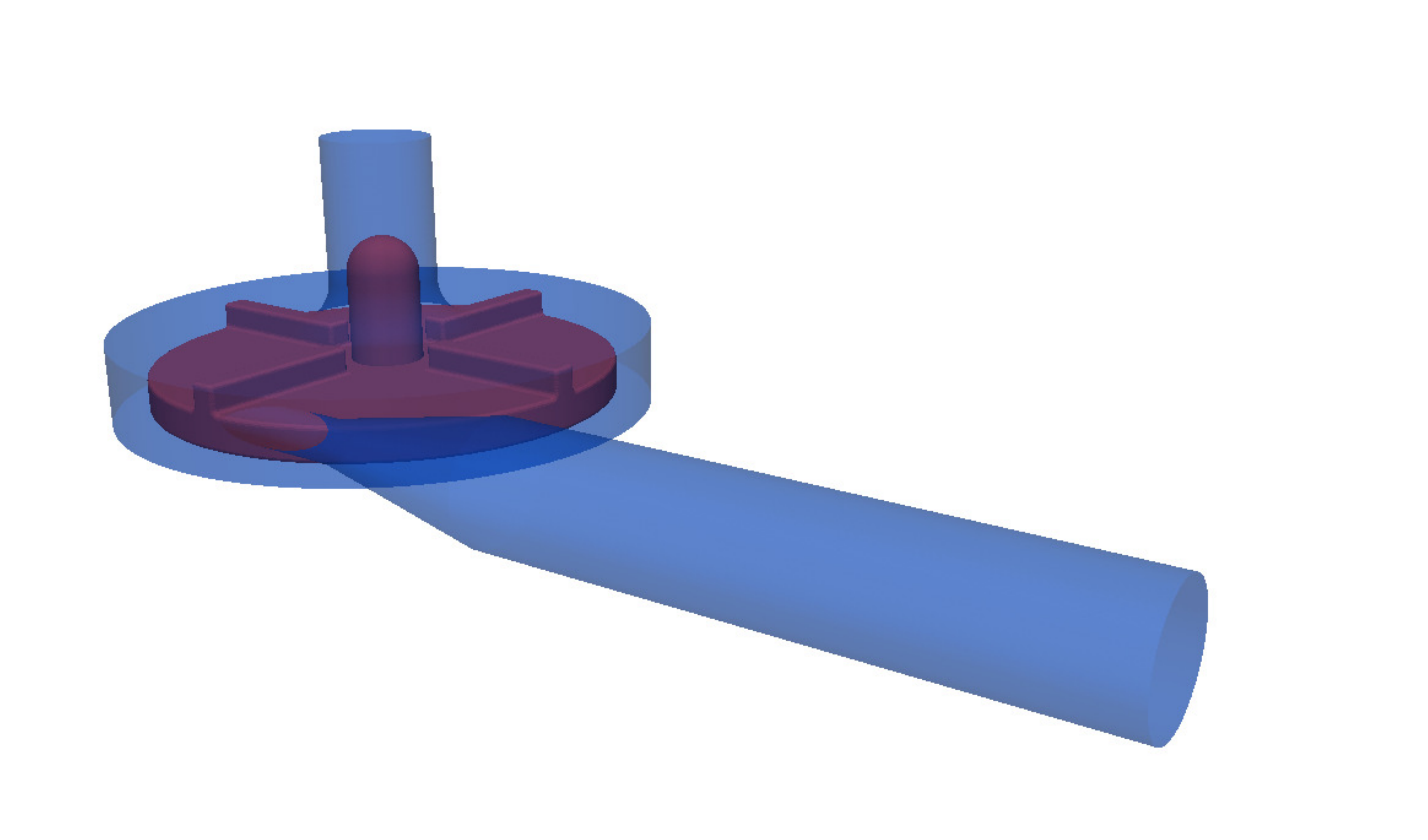}};
	
	\coordinate (pt1) at (3.0,4.6);
	\coordinate (pt2) at (2.1,4.1);
	\coordinate (pt3) at (8,1.5);
	\coordinate (pt4) at (6.8,1);
	\coordinate (pt5) at (1,3.5);
	\coordinate (pt6) at (1.7,2.6);
	
	\coordinate (pt7) at (2.1,0.5);
	\coordinate (pt8) at (2.1,1.8);
	
	\coordinate (pt9) at (1.8,1);
	\coordinate (pt10) at (2.5,0.9);
	
	\draw[->, line width=1.0pt] (pt1) to node[pos=-0.3] {\small $\Gamma_{in}$} (pt2);
	\draw[->, line width=1.0pt] (pt3) to node[pos=-0.3] {\small $\Gamma_{out}$} (pt4);
	\draw[->, line width=1.0pt] (pt5) to node[pos=-0.3] {\small $\Gamma_{rotor}$} (pt6);

	\draw[dashed, line width=1.0pt] (pt7) to (pt8);
	\draw[->, line width=1.0pt] (pt9) to [bend right=50] node[pos=1.2] {\small $\omega$} (pt10);
	
	\end{tikzpicture}
	\caption{Illustration of the boundaries and the axis of rotation on the blood pump geometry.}
	\label{fig:pump_geom}
\end{figure}

We prescribe a "do-nothing" condition on the outflow boundary $\Gamma_{out}$, two different Dirichlet boundary conditions are applied on the inflow boundary $\Gamma_{in}$ and the rotor's surface $\Gamma_{rotor}$ separately. The rest of boundary is covered by "no-slip" condition (\Cref{fig:pump_geom}). The inflow boundary condition is modeled with a Poiseuille profile:

\begin{equation}
	\bm{g} = \begin{bmatrix}
	0\\
	0\\
	-U_{max} (1 - l^2/L^2)
	\end{bmatrix} 
	\;,
\end{equation}

\noindent $L$ is the radius of the circular inflow boundary, $l$ is the distance from a point on $\Gamma_{in}$ to the center point. $U_{max}$ is the maximum inflow speed, and $U_{max} > 0$, as the inflow direction is in $-\bm{e}_3 = [0,0,-1]^T$.

The angular speed is defined as:

\begin{equation}
\bm{\omega} = \begin{bmatrix}
0\\
0\\
\omega
\end{bmatrix}
\;,
\end{equation}
\noindent the direction of the axis of rotation for the rotor is showed in \Cref{fig:pump_geom}.

\subsection{Shear layer update approach}

With the purpose of modeling a rotating machinery, we propose here a shear layer update approach, which is inherited from SSMUM \cite{Behr1999a, Behr2003a}, we employ though two layers instead of one to achieve the efficiency of the update process once a regeneration of mesh is required. In general, for the mesh patching technique \cite{Steger1983a}, the computational domain is divided by two, static and rotating domain ($\Omega_{stat}$ and $\Omega_{rot}$). The static domain stays unchanged, the rotating domain evolves corresponding the speed of blade.

\begin{figure}[t]
	\centering
	\begin{subfigure}[b]{0.475\textwidth}
		\centering
		\begin{tikzpicture}
		\node[anchor=south west,inner sep=0] (image) at (0,0)
		{\includegraphics[height=0.7\textwidth]{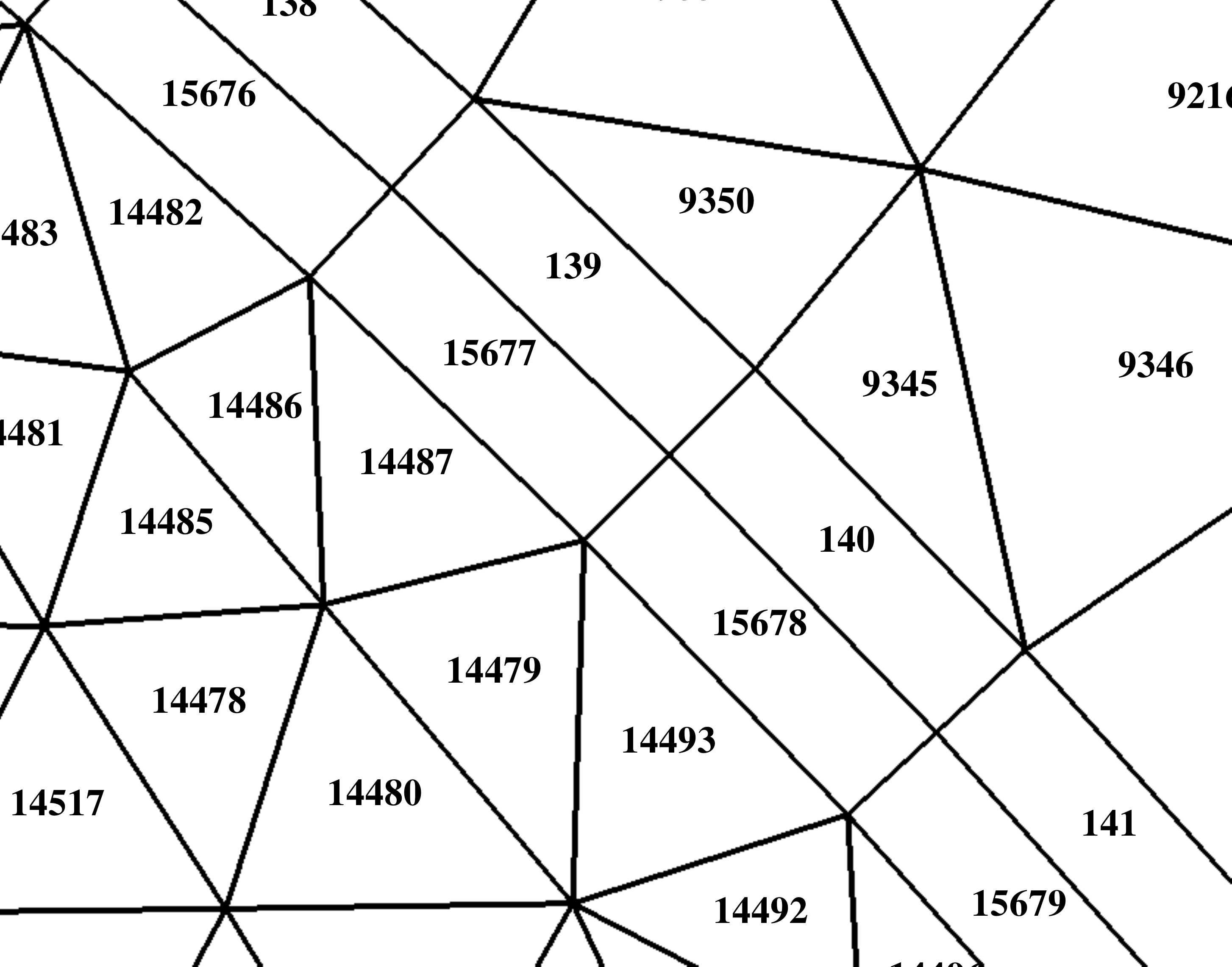}};
		
		\coordinate (pt1) at (2.5,1.5);
		\coordinate (pt2) at (1.5,2.5);
		\draw[->, line width=1.5pt,red] (pt1) to [bend right=10] (pt2);
		
		\end{tikzpicture}
		\caption{}
	\end{subfigure}
	~
	\begin{subfigure}[b]{0.475\textwidth}
		\centering
		\begin{tikzpicture}
		\node[anchor=south west,inner sep=0] (image) at (0,0)
		{\includegraphics[height=0.7\textwidth]{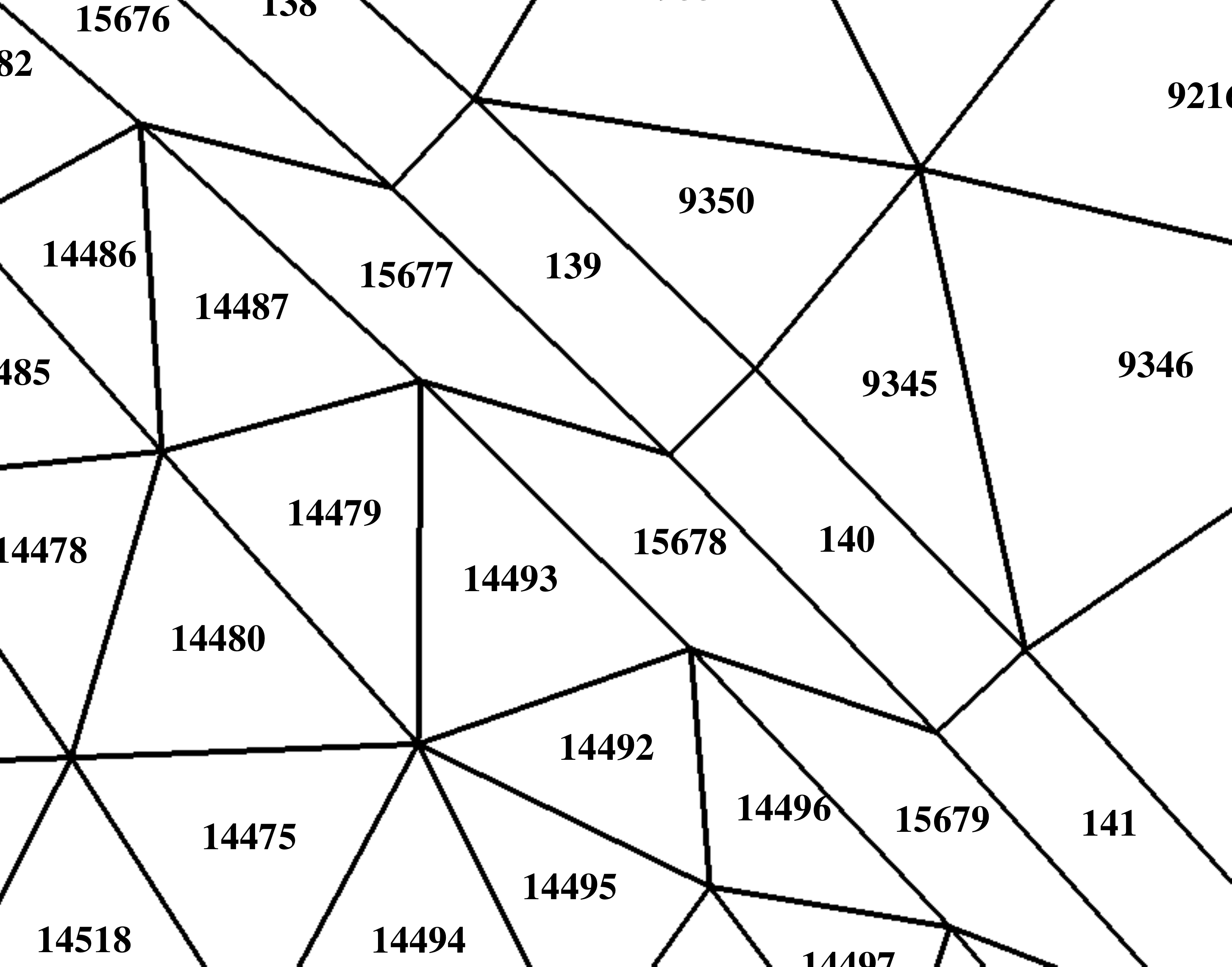}};
		
		\coordinate (pt1) at (2.5,1.5);
		\coordinate (pt2) at (1.5,2.5);
		\draw[->, line width=1.5pt,red] (pt1) to [bend right=10] (pt2);
		
		\end{tikzpicture}
		\caption{}
	\end{subfigure}
	\\
	\begin{subfigure}[b]{0.475\textwidth}
		\centering
		\begin{tikzpicture}
		\node[anchor=south west,inner sep=0] (image) at (0,0)
		{\includegraphics[height=0.7\textwidth]{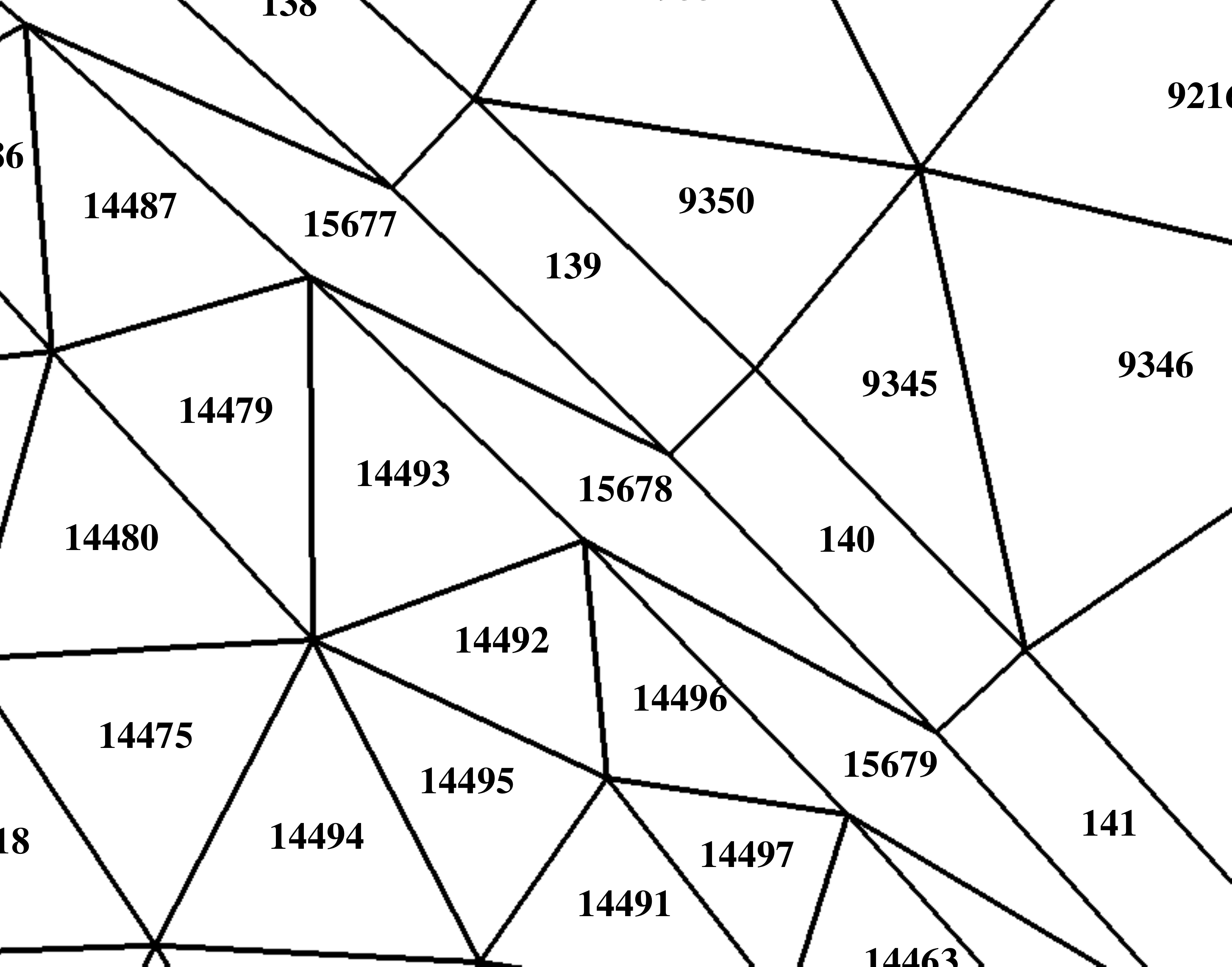}};
		
		\coordinate (pt1) at (2.5,1.5);
		\coordinate (pt2) at (1.5,2.5);
		\draw[->, line width=1.5pt,red] (pt1) to [bend right=10] (pt2);
		
		\end{tikzpicture}
		\caption{}
	\end{subfigure}
	~
	\begin{subfigure}[b]{0.475\textwidth}
		\centering
		\begin{tikzpicture}
		\node[anchor=south west,inner sep=0] (image) at (0,0)
		{\includegraphics[height=0.7\textwidth]{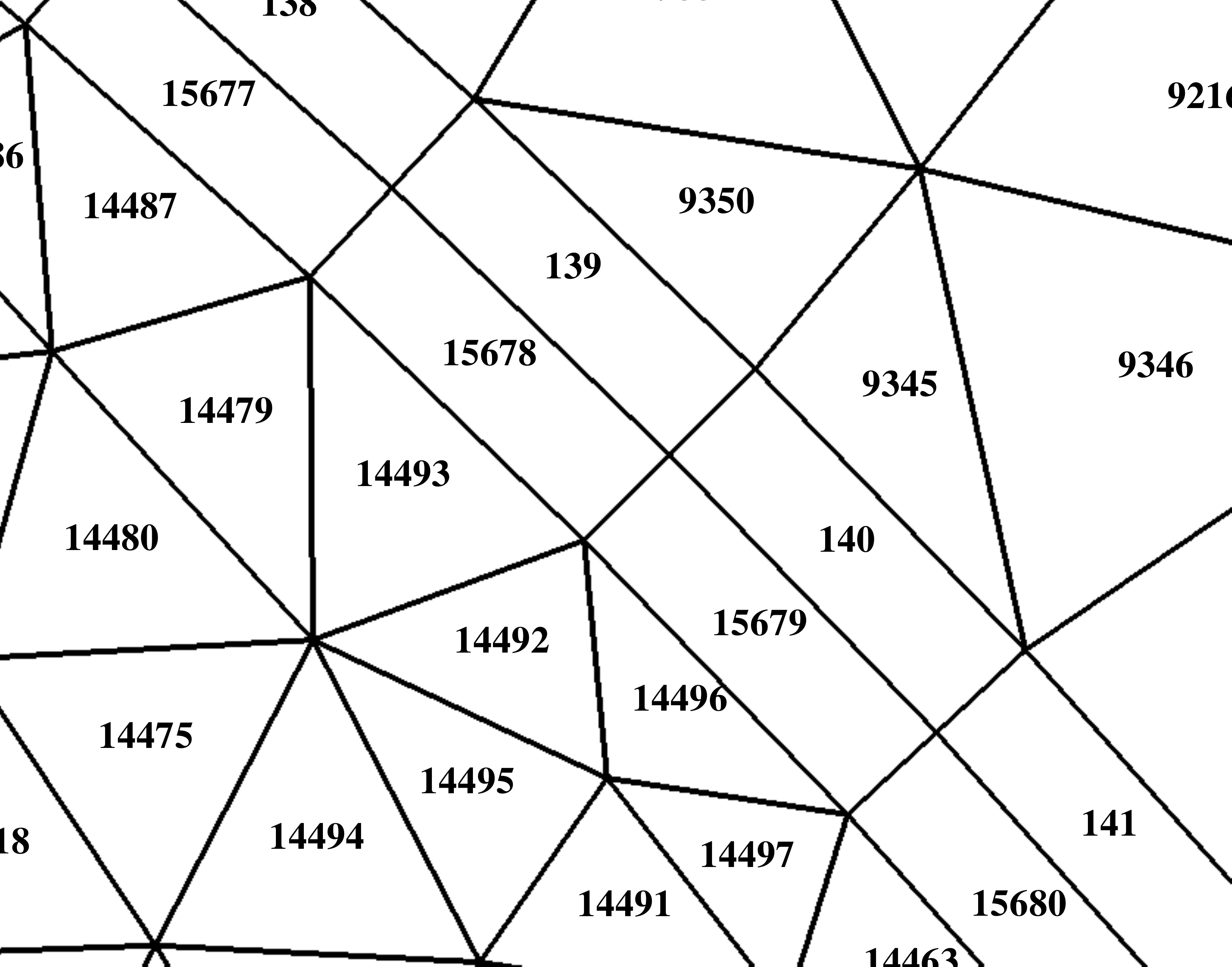}};
		
		\coordinate (pt1) at (2.5,1.5);
		\coordinate (pt2) at (1.5,2.5);
		\draw[->, line width=1.5pt,red] (pt1) to [bend right=10] (pt2);
		
		\end{tikzpicture}
		\caption{}
	\end{subfigure}
	\caption{Illustration of mesh motion during shear process and reconnecting process.}
	\label{fig:shear_demo}
\end{figure}

\Cref{fig:shear_demo} demonstrates the main procedure of moving mesh for rotating machinery modeling, static and rotating domain own one layer separately and each cell on those two layers has the same shape. The border between these two domains is therefore the interface between the two identical layers as well. At the beginning, the initial position of the mesh looks like in step a, when the rotating part moves in relation to the rotor, we first shear one of these layers in order to obtain more freedom of movement, as we are limited by the mesh size most of the time. The connectivity of mesh cell remains unchanged during the shear process (step b), only spatial location needs to be modified, the adaptation is then uncomplicated. When the shearing attends the same length as the cell (step c). Afterwards, the vertices on the common interface detach, and reconnect to the subsequent points (step d).

If we could ensure that the local mesh only contains either a part of static mesh or rotating mesh (it can be accomplished easily by using group of communicator or similar concept of parallel communication), thus, we would benefit from this setting, as the structure of the local vector does not alter \cite{hiflow2013a}. Therefore, only values on the ghost cells in the local vector need to be updated. For that reason, projecting solution into a new linear algebra structure is avoided, the update process after reconnecting vertices can be accelerated.

\subsection{Variational Multiscale method (VMS)} 

The Reynolds number for a rotating machinery is defined like:

\begin{equation}
Re := \frac{\omega D^2 \rho}{\mu} \;,
\label{eq:reynolds}
\end{equation}

\noindent due to the high rotation speed $\omega$ of our instrument (2500 RPM \footnote{RPM: revolutions per minute.}), the Reynolds number is approximately $21000$ in our blood pump, an additional high Reynolds number flow modeling is required, we choose in this work the Variational Multiscale method (VMS) for this purpose. This technique considers a separation of complete solution scale into two different groups: resolvable scale and unresolvable scale. In general, VMS can be considered as a kind of stabilized finite element technique \cite{Tezduyar1991a}, whereas, it can also be meanwhile interpreted as large eddy simulation (LES) for turbulence modeling. Without describing the theory, we illustrate only the projection-based finite element variational multiscale method for Navier-Stokes equations. Further insight about this method can be found in \cite{Hughes1998a, Hughes2000a, John2005a, Gravemeier2007a}.

The variational multiscale formulation can only be seen in a weak form:

\begin{subequations}
\begin{align}
\begin{split}
\int_{\Omega} (\frac{\partial \bm{u}}{\partial t} + ((\bm{u} - \bm{u}^r) \cdot \nabla) \bm{u}) \cdot \bm{v} \: d \Omega + \int_{\Omega} \frac{\mu}{\rho} \nabla{\bm{u}} :\nabla \bm{v} \: d\Omega - \int_{\Omega} \frac{1}{\rho} p \nabla \cdot \bm{v} \: d \Omega \: d \Omega&\\
+ \int_{\Omega} [\frac{\partial \bm{u}}{\partial t} + ((\bm{u} -\bm{u}^r) \cdot \nabla) \bm{u} - \frac{\mu}{\rho} \Delta \bm{u} + \frac{1}{\rho} \nabla p] \tau_{M} ((\bm{u} - \bm{u}^r) \cdot \nabla \bm{v}) \: d \Omega&\\
+ \int_{\Omega} (\nabla \cdot \bm{u}) \cdot \tau_{C} (\nabla \cdot \bm{v}) \: d \Omega & = 0 \;,
\end{split}\\
\begin{split}
\int_{\Omega}q \nabla \cdot \bm{u} \: d \Omega&\\
+ \int_{\Omega} \tau_{M} \nabla q [\frac{\partial \bm{u}}{\partial t} + (\bm{u} \cdot \nabla) \bm{u} - \frac{\mu}{\rho} \Delta \bm{u} + \frac{1}{\rho} \nabla p] \: d \Omega & = 0 \;.
\end{split}
\end{align}
\label{eq:vms}
\end{subequations}

\noindent $\bm{v}$ and $q$ are the test functions for velocity and pressure. $\tau_M$ and $\tau_C$ are the elementwise constant stabilization operators, which project the unsolvable solution on the fine scale into the coarse scale based the residual of the governing equations.

\subsection{Uncertainty model}
\label{sec2:subsec3:uq_model}

We consider three different parametric uncertainties: the inflow boundary condition $\bm{g}$, the angular speed $\bm{\omega}$ and the dynamic viscosity $\mu$. We model each of those uncertain parameters with independent, uniformly distributed random variables $\xi_i \sim U(-1,1), i = 1,2,3$,  it reads:

\begin{subequations}
\begin{align}
\bm{g} &= \bm{g}_0 + \bm{g}_1 \xi_1 \;, \\
\bm{\omega} &= \bm{\omega}_0 + \bm{\omega}_2 \xi_2 \;, \\
\mu &= \mu_0 + \mu_3 \xi_3 \;.
\end{align}
\label{eq:uq_input}
\end{subequations}

\noindent Where, $\bm{g}_1 = \sigma_1 \bm{g}_0$, $\bm{\omega}_2 = \sigma_2 \bm{\omega}_0$ and $\mu_3 = \sigma_3 \mu_0$. $\sigma_i$ are the decay factors respect to the mean value, thus $0 < \sigma_i < 1$. With reference to PCE, we define a multivariate random variable $\bm{\xi} := (\xi_1, \xi_2, \xi_3)$. Accordingly, $\bm{\xi}$ allows us directly map the outcomes of an abstract probability space $(\Omega, \mathcal{A}, \mathbb{P})$ to a subset $T$ of $\mathbb{R}^3$. Afterwards, we can express our stochastic solution immediately with the aid of $\bm{\xi}$.

\subsection{Stochastic Galerkin projection}

The velocity and pressure in \Cref{eq:ns} are expressed with the Polynomial Chaos Expansion (PCE) method \cite{LeMaitre2010a} by employing the orthogonal polynomial $\psi$:

\begin{subequations}
\begin{align}
\bm{u}(\bm{x}, \bm{\xi}) &= \sum_{i=0}^{\infty} \bm{u}_i \psi_i(\bm{\xi}) \;,\\
p(\bm{x}, \bm{\xi}) &= \sum_{i=0}^{\infty} p_i \psi_i(\bm{\xi}) \;.
\end{align}
\label{eq:pce_inf}
\end{subequations}

\noindent $\psi_i$ represents the Chaos Polynomials, more specifically in this paper, their are the normalized Legendre Polynomials. $\bm{u}$ and $p$ follow certain general assumptions, such as square-integrable respect to $\bm{\xi}$, more details can be found in \cite{LeMaitre2010a}. The orthogonality of $\psi_i$ with respect to the probability density function of $\bm{\xi}$ can be explicit written as:

\begin{equation}
\int_{[-1,1]^3} \psi_i(\bm{\xi}) \psi_h(\bm{\xi}) \frac{1}{2^3} d\bm{\xi} = \delta_{ij} \;,
\end{equation}

\noindent $\delta_{ij}$ here is the Kronecker delta function. In order to be able of computing our stochastic solution numerically, we need to truncate \Cref{eq:pce_inf} up to certain polynomials order $No$, it yields:

\begin{subequations}
\begin{align}
\bm{u}(\bm{x}, \bm{\xi}) &\approx \sum_{i=0}^{P} \bm{u}_i \psi_i(\bm{\xi}) \;,\\
p(\bm{x}, \bm{\xi}) &\approx \sum_{i=0}^{P} p_i \psi_i(\bm{\xi}) \;.
\end{align}
\label{eq:truncation}
\end{subequations}

\noindent $P + 1 = (M+No)!/(M!No!)$ is the total number of the PC modes, $M$ is the number of uncertain variables, in our case, $M = 3$.

Therefore, we could pursuit the generalized Polynomial Chaos Expansion (gPCE) \cite{Xiu2002a} procedure by first inserting \Cref{eq:truncation} into \Cref{eq:vms}, then multiplying an additional polynomial $\psi_k, k = 0,...,P$ on both side of governing equations, and taking $L^2$ inner product on $L^2(T)$. By considering the orthogonality of the polynomials, the regarding momentum equation and mass conservation can be thus written as:

\begin{subequations}
\begin{align}
\frac{\partial \bm{u}_k}{\partial t} \bm{v}_k + \sum_{i=0}^{P} \sum_{j=0}^{P} ((\bm{u}_i - \bm{u}_i^r) \cdot \nabla) \bm{u}_j \bm{v}_k c_{ijk} + \sum_{i=0}^{P} \sum_{j=0}^{P} \frac{\mu_i}{\rho} \nabla \bm{u}_i : \nabla \bm{v}_j c_{ijk} + \frac{1}{\rho} p_i \nabla \cdot \bm{v}_k \\ \nonumber
+ \tau_M ((\bm{u}_k - \bm{u}_k^r) \cdot \nabla \bm{v}_k) [\frac{\partial \bm{u}_k}{\partial t} + \sum_{i=0}^{P} \sum_{j=0}^{P} ((\bm{u}_i - \bm{u}_j^r) \cdot \nabla) \bm{u}_j c_{ijk} \\ \nonumber
- \sum_{i=0}^{P} \sum_{j=0}^{P} \frac{\mu_i}{\rho} \Delta \bm{u}_j c_{ijk} + \frac{1}{\rho} \nabla p_k] + (\nabla \cdot \bm{u}_k) \tau_C (\nabla \cdot \bm{v}_k) \;, &&\text{in} \: \Omega\;, \\
q_k \nabla \cdot \bm{u}_k \\ \nonumber
\tau_M \nabla q_k [\frac{\partial \bm{u}_k}{\partial t} + \sum_{i=0}^{P} \sum_{j=0}^{P} ((\bm{u}_i - \bm{u}_i^r) \cdot \nabla) \bm{u}_j c_{ijk} \\ \nonumber
- \sum_{i=0}^{P} \sum_{j=0}^{P} \frac{\mu_i}{\rho} \Delta \bm{u}_j c_{ijk} + \frac{1}{\rho} \nabla p_k] \;, &&\text{in} \: \Omega \;.
\end{align}
\text{for $k = 0, ..., P$, and $c_{ijk} := <\psi_i \psi_j, \psi_k>$.}
\label{eq:ns_pce}
\end{subequations}

So fair, we introduce the variational multiscale formulation for the incompressible Navier-Stokes equations in a rotating machinery based on our moving mesh technique, the modeling of input uncertain parameters and the generalized Polynomial Chaos Expansion according to considered problem is also presented. In the following section, we will focus on the numerical methods for solving this specially system.

\section{NUMERICAL METHODS}
\label{sec:num_method}

The nonlinear terms in \Cref{eq:ns_pce} are linearized with Newton method, the general scheme can be written as:

\begin{equation}
\mathbf{J}_{\mathbf{F}} (\bm{x}_n) (\bm{x}_{n+1} - \bm{x}_n) = \mathbf{F}(\bm{x}_n) \;,
\label{eq:newton}
\end{equation}

\noindent $\mathbf{F}$ is the system equation, $\mathbf{J}_F$ is the Jacobian matrix of $\mathbf{F}$. $\bm{x}_n$ is the solution vector at $n$-th Newton iteration, $\bm{x}_n = [\bm{u}_n, p_n]^T$. At each Newton iteration, an update ($\bm{x}_{n+1} - 
\bm{x}_n$) of the solution vector will be computed, until it meets to certain convergence criterion.

The spatial part of our problem is therefore discretized by the finite element method with equal order P1/P1 element for velocity and pressure respectively, thanks to VMS, no Taylor-Hood element is necessary. Afterwards, we will obtain a linear system of equations for the variational multiscale formulation for the Navier-Stokes equations in rotating system. The stiffness matrix $A(\bm{x}_n) \in \mathbb{R}^{(P+1)N, (P+1)N}$ can be expressed as:

\begin{equation}
A(\bm{x_n}) = \sum_{i=0}^{P} K_i \otimes A_i(\bm{x_n}) \;.
\label{eq:stiffness}
\end{equation}

\noindent Note that, $N$ is the number of degree of freedom for the finite element discretization, $\bm{x}_n$ is the linearized point. $A_i(\bm{x}_n) \in \mathbb{R}^{N,N}, i=0, ..., P$ are the Kronecker factors and $K_i \in \mathbb{R}^{P+1,P+1}, i=0, ..., P$ denote the stochastic Galerkin matrices defined by $(K_i)_{j,k} = c_{ijk}$, which will play an important role in the multilevel method.

\subsection{Multilevel method}

Solving a such complex stochastic Galerkin system is very challenging, choosing a good solving strategy is very crucial. The multilevel method has been applied with good convergence behavior \cite{Schick2015a, Eveline2010a, Schick2014a}.

The stochastic Galerkin system possesses a hierarchical structure due to the polynomial construction, thus we make use of this feature. If space $\mathcal{S}_l$ is spanned by the Chaos Polynomials:

\begin{equation}
\mathcal{S}_l = \text{span} \{\psi_0, ..., \psi_{P_l}\} \;,
\label{eq:space_spanned}
\end{equation}

\noindent $l \in \mathbb{N}, \; l \leq No$ is certain polynomial degree, $P_l = (M+l)!/(M!l!)$. $M$ is again the number of uncertain parameters, in this paper, $M=3$. $P_l$ is the total number of PC mode with respect to the polynomial order $l$. The hierarchical space $\mathcal{S}$ can be also expressed as a nested sequence of spaces with the consideration of the total polynomial degree $No$:

\begin{equation}
\mathcal{S}_0 \subseteq \mathcal{S}_1 \subseteq \cdots \subseteq \mathcal{S}_l \cdots \subseteq \mathcal{S}_{No} \;.
\label{eq:pce_subspace}
\end{equation}

The discretized Newton step of \Cref{eq:ns_pce} at iteration $n$ based on \Cref{eq:space_spanned} can be found like:

\begin{equation}
\sum_{i=0}^{P_{No}} K_i \otimes A_i(\bm{x}_n) \bm{\tilde{x}}_n^{No} = - \bm{b}_n^{No} \;.
\end{equation}

\noindent $\bm{\tilde{x}}_n^{No}, \bm{b}_n^{No} \in \mathbb{R}^{N(P+1)}$, and $\bm{\tilde{x}}_n^{No} = \bm{x}_{n+1}^{No} - \bm{x}_{n}^{No}$. We bring into play the principle of multigrid method \cite{Brandt1973a, Brandt1977a, Hackbusch1979a}, and consider the polynomial degree $l$ in \Cref{eq:pce_subspace} as the "grid level". Hence, the restriction operator $\mathcal{R}_{l-1}$ and the prolongation operator $\mathcal{P}_l$ are naturally defined as a $L^2$ projection from $\mathcal{S}_{l-1}$ to $\mathcal{S}_l$ or vice versa.

One of very crucial procedure for Multilevel method (as well as for Multigrid scheme) is to choose an appropriate smoother, we utilize here the Mean based preconditioner as our smoother \cite{Schick2015a}, we apply $\theta$ times smoothing process to a given initial solution $x_0^l$:

\begin{equation}
x_{k+1}^l := \mathcal{B}_l x_l^l = x_k^l + (I_d \otimes A_0)^{-1} \;.
\label{eq:smoothing}
\end{equation}

\begin{figure}[t]
	\begin{algorithm}[H]
		\begin{algorithmic}[1]
			\IF{$l=0$}
			\STATE solve $A_0 x^0 = b^0$
			\ELSE
			\STATE $x^l = \mathcal{S}^l x^l$ \quad ($\nu_1$ times pre-smoothing)
			\STATE $r^l = b^l - \sum_{i=0}^{P_l} (K_i \otimes A_i) x^l$ \quad (compute residual)
			\STATE $r^{l-1} = \mathcal{R}_{l-1}r^l$ \quad (restriction)
			\FOR{$i=1$ to $\mu$}
			\STATE ML($b^{l-1}$, $x^{l-1}$, $l$) \quad (V-cycle/W-cycle)
			\ENDFOR
			\STATE $c^l = \mathcal{P}_l c^{l-1}$ \quad (prolongation)
			\STATE $x^l = x^l + c^l$ \quad (update with correction)
			\STATE $x^l = \mathcal{S}^l x^l$ \quad ($\nu_2$ times post-smoothing)
			\ENDIF
		\end{algorithmic}
		\caption{PCE Multilevel preconditioner/solver : ML}
		\label{alg:ml}
	\end{algorithm}
	\caption{One cycle of the multilevel method: ML($x^l,b^l,l)$ given a vector $x^l$ and right hand side $b^l$ on level $l$. $\mu = 1$ results in a $V$-cycle, $\mu=2$ in a $W$-cycle.}
\end{figure}

\noindent Note that, the subscript $k$ indicates the iteration number within the smoothing procedure, which is not the same for Newton step in \Cref{eq:newton} and \Cref{eq:stiffness}. A pseudo algorithm about our stochastic Multilevel solver/preconditioner can be found in \Cref{alg:ml}.

\section{NUMERICAL RESULTS}
\label{sec:num_result}

We use the Crank-Nicolson time stepping scheme for the instationary Navier-Stokes equations, and the nonlinear system \Cref{eq:ns_pce} is solved via the inexact Newton scheme, it implies that only an approximated solution is applied for each Newton step. We applied the strategy "choice 1" of Eisenstat and Walker in \cite{Eisenstat1996a} with an initial forcing term equals to $0.5$.

\begin{table}[h]
	\centering
	\begin{tabular}{|l|r|l|r|}
		\cline{1-4}
		Inflow maximal speed ($m/s$) & 0.5 & Inflow speed variation ($\sigma_1$) & 10\%\\ \cline{1-4}
		Dynamic viscosity ($N\cdot s/m^{2}$) & 0.0035 & Viscosity variation ($\sigma_3$) & 10\% \\ \cline{1-4}
		Angular speed ($rad/s$) & 261.8 & Angular speed variation ($\sigma_2$) & 10\% \\ \cline{1-4}
		RPM & 2500 & Density ($Kg/m^{3}$) & 1035  \\ \cline{1-4}
	\end{tabular}
	\caption{Model parameter values.}
	\label{tab:input_paramter}
\end{table}

For each Newton step, we apply an iterative Krylov subspace solver, the flexible generalized minimal residual method (FGMRES), with a Multilevel method preconditioner, which is presented in previous section. For the mean block solver, we use the generalized minimal residual method (GMRES), which is preconditioned with Schur Complement preconditioner. We use an inexact Multilevel method here, it indicates that we compute the inverse of $A_0$ only respect to certain lower accuracy criterion. In this case, we solve the linear system with $A_0$ (in \Cref{eq:smoothing}) only up to a relative error $1.0e-1$. In contract, the relative accuracy of the Newton step is set to $1e-9$.

\Cref{tab:input_paramter} indicates the information about input parameters, the Reynolds number, according to \Cref{eq:reynolds} is about $210000$. The geometry discretization contains $2,984,259$ unstructured cells, it results thus $2,274,904$ degrees of freedom for the deterministic case. As mentioned in \Cref{sec2:subsec3:uq_model}, we consider three different input uncertain parameters:inflow boundary condition, dynamic viscosity and angular speed of the rotor. In this work, we consider the situation that the polynomial degree equals to $3$, which gives us in total $20$ PC modes. Therefore, the total number of degrees of freedom is about $45.5$ Millions.

\begin{figure}
	\centering
	\includegraphics[width=0.8\textwidth]{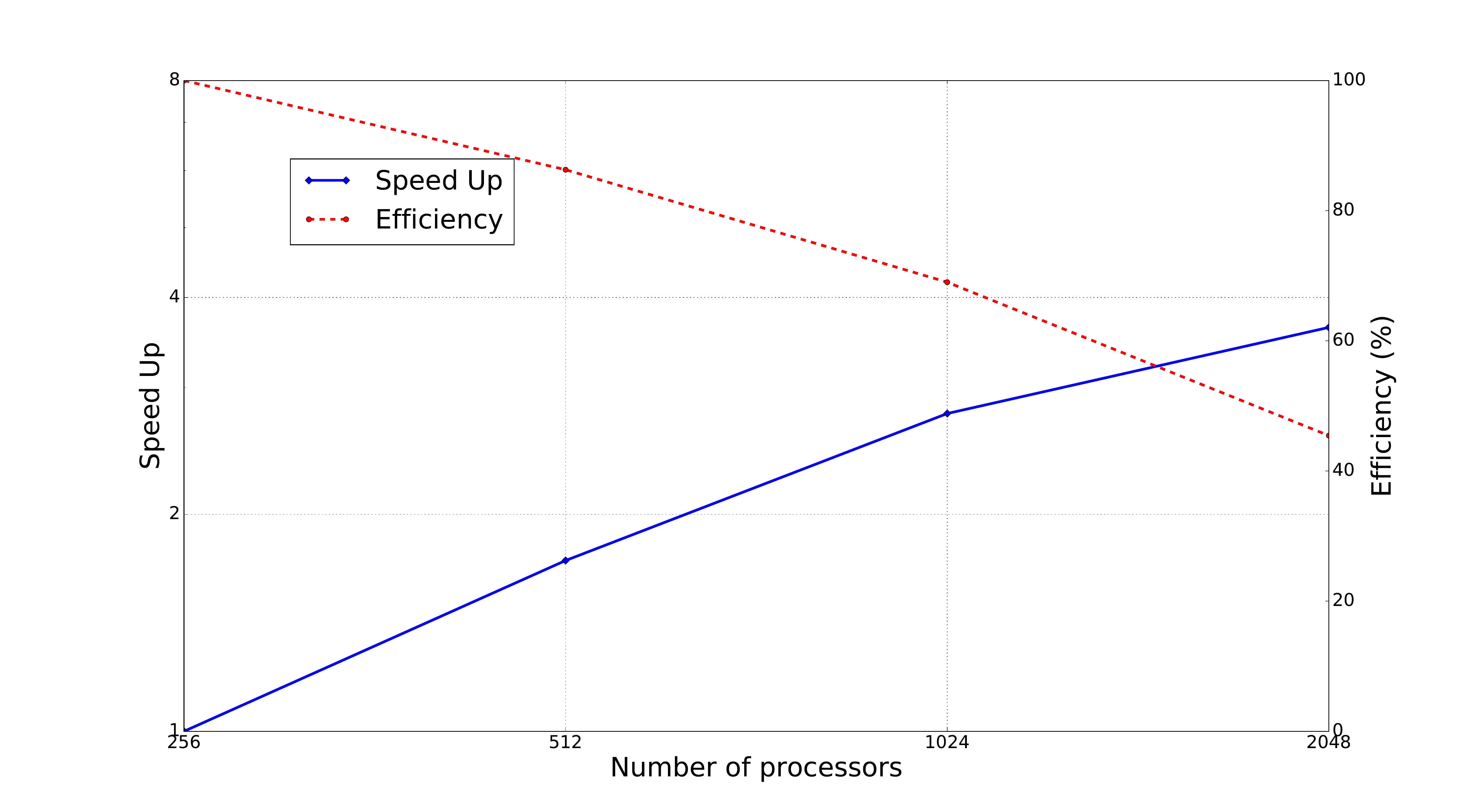}
	\caption{Scalability test with the first $50$ time steps, $3$ uncertainties parameters and polynomial degrees equals to $2$.}
	\label{fig:scalability}
\end{figure}

\Cref{fig:scalability} shows a scalability test by using the information of the first $50$ time steps. The speed up and the efficiency is defined as $S_p:= \frac{T_1}{T_p}, E_p := \frac{S_p}{p}$. We find out that the efficiency of our full solving process could still catch around $48\%$ when the number of processor is up to $2048$. We have to mention that we start our test with $256$ processors, because it is the minimal requirement of cores we can use regarding to our problem size, the efficiency of the Multilevel preconditioner itself can perform better if we consider a simpler problem and start the scalability with only one processor.

\begin{figure}
	\centering
	\begin{subfigure}[b]{0.45\textwidth}
	\includegraphics[width=\textwidth]{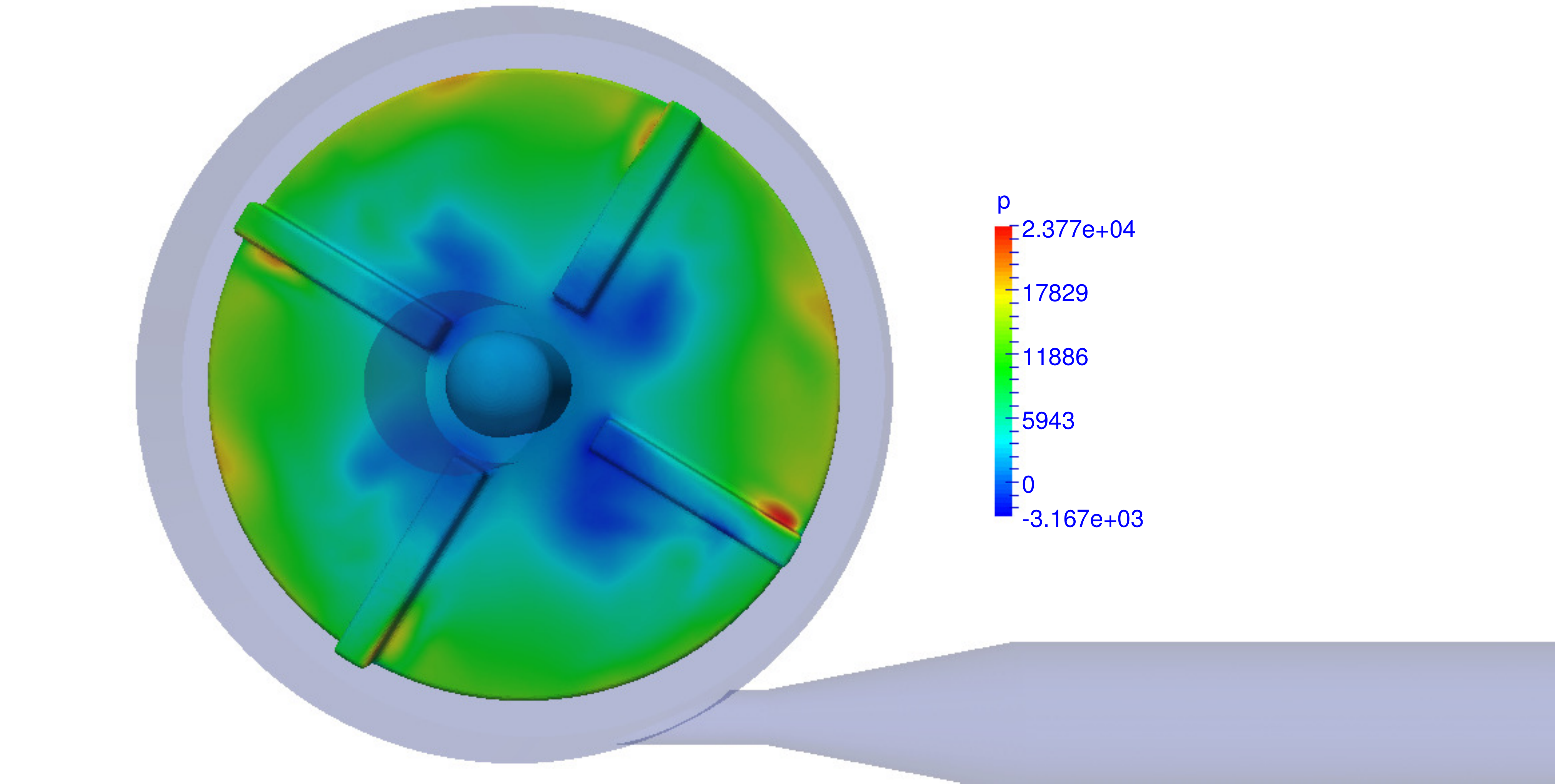}
	\caption{Mean value.}
	\end{subfigure}
	~
	\centering
	\begin{subfigure}[b]{0.45\textwidth}
		\includegraphics[width=\textwidth]{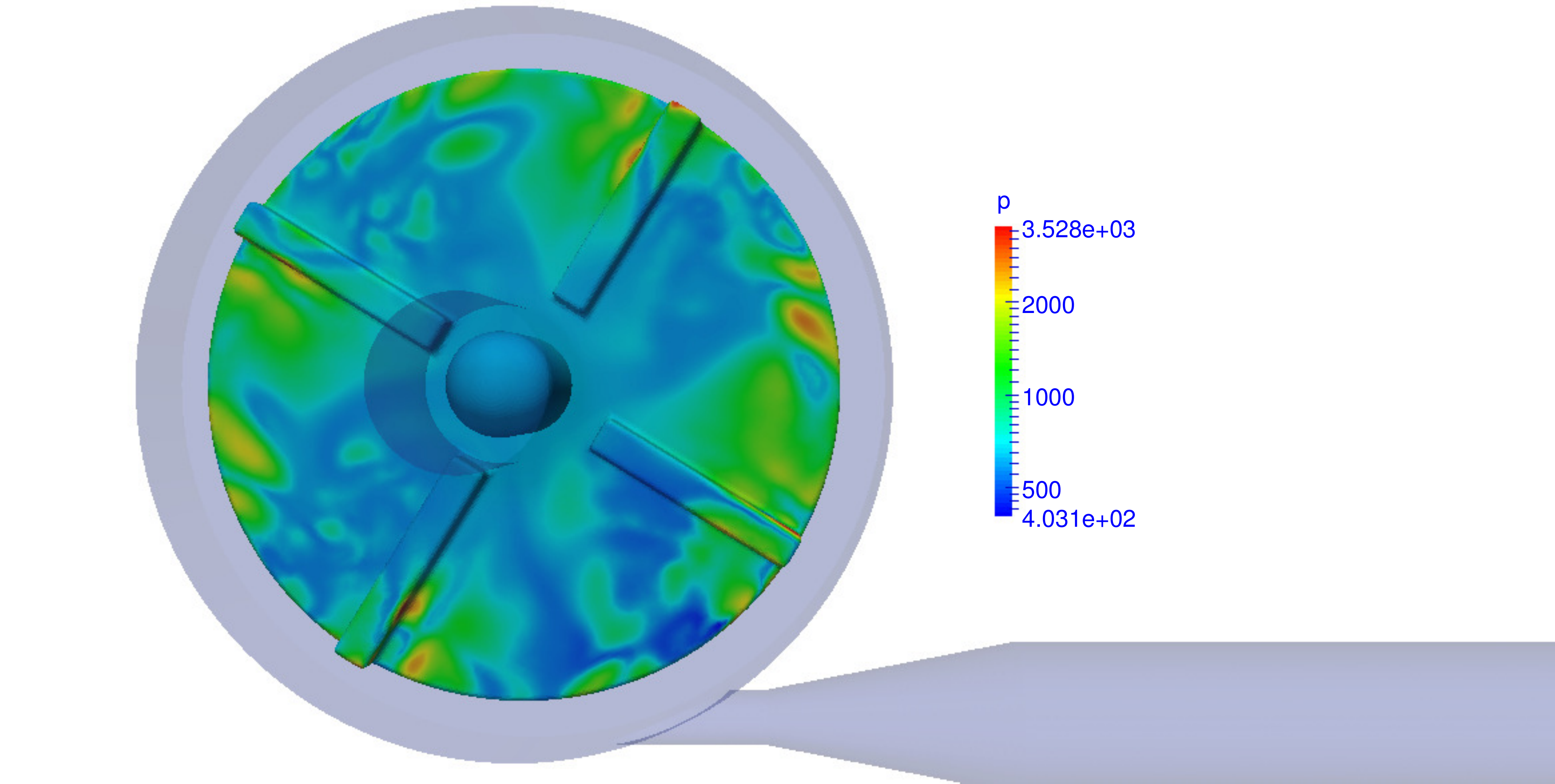}
		\caption{Standard deviation.}
	\end{subfigure}
	\caption{The mean value and standard deviation of the pressure at time step = 500.}
	\label{fig:uq_pressure}
\end{figure}

\begin{figure}
	\centering
	\begin{subfigure}[b]{0.45\textwidth}
		\includegraphics[width=\textwidth]{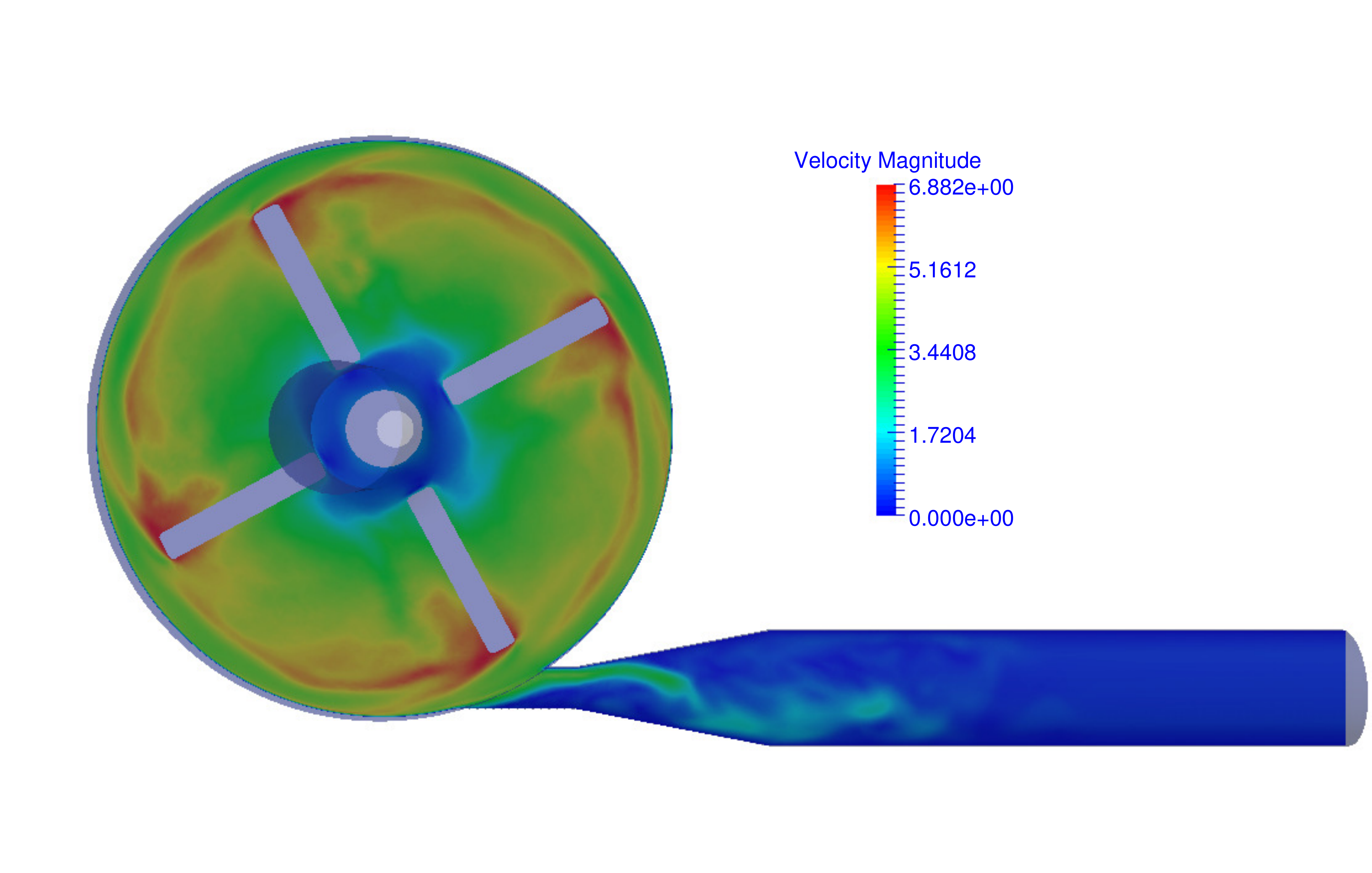}
		\caption{Mean value.}
	\end{subfigure}
	~
	\centering
	\begin{subfigure}[b]{0.45\textwidth}
		\includegraphics[width=\textwidth]{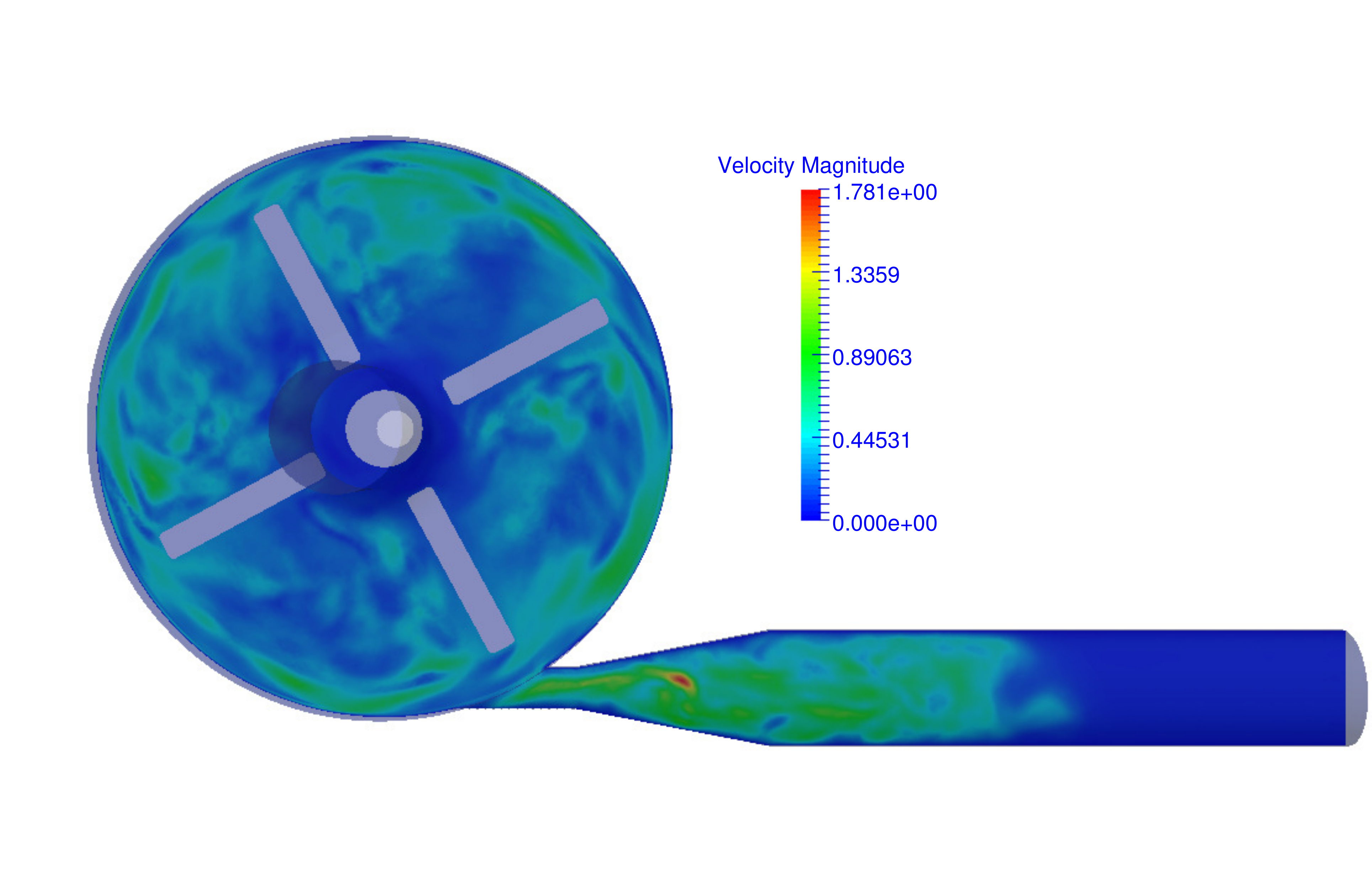}
		\caption{Standard deviation.}
	\end{subfigure}
	\caption{The mean value and standard deviation of the velocity at time step = 500.}
	\label{fig:uq_velocity}
\end{figure}

\begin{figure}[t]
	\centering
	\includegraphics[width=0.8\textwidth]{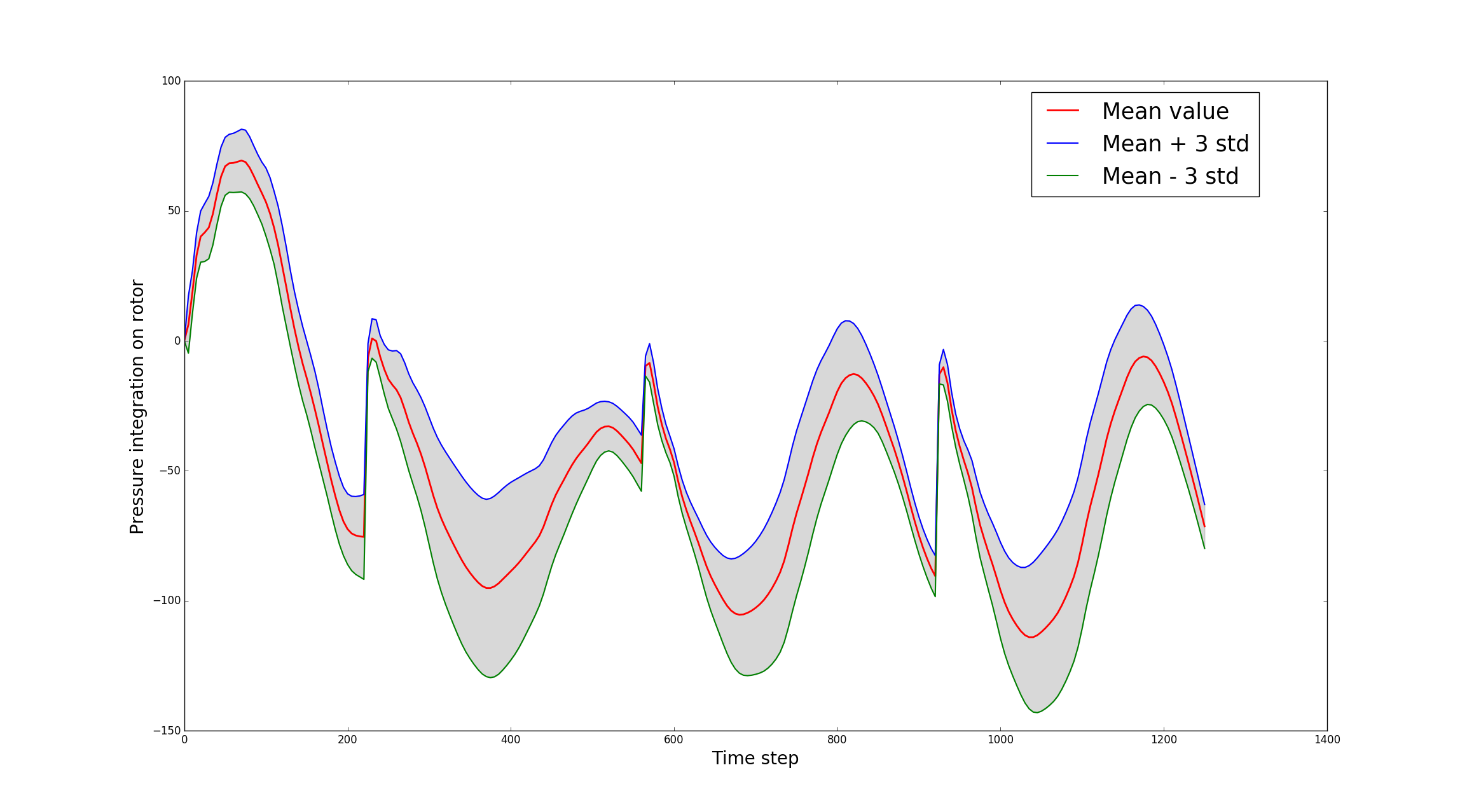}
	\caption{Mean value and mean $\pm$ 3 standard deviation for the pressure integration over the rotor.}
	\label{fig:pressure_integrate}
\end{figure}

\Cref{fig:uq_pressure} shows the mean value and standard deviation of pressure distribution on the rotor, the standard deviation follows slightly the magnitude of the mean value, as at the center it is lower and on the hub of the blade is higher. But the uncertainty distribution arises also at certain locations where the pressure is less important. \Cref{fig:uq_velocity} represents the mean value and the standard deviation of the velocity, the standard deviation becomes higher after the flow at the outlet, it might be due to the acceleration after the nozzle structure. \Cref{fig:pressure_integrate} represents the value of pressure integration over the rotor. It shows the mean value (red) and mean $\pm$ $3$ standard deviation involving along time.

\section{CONCLUSIONS}
\label{sec:conclusion}

This work is a continuation of our last contribution \cite{Schick2015a}, which considered a simplified geometry and an artificial stabilized laminar flow. Within this work, we introduce first with our moving mesh strategy in order to be able of modeling this rotating machine in unsteady state, the evolution speed is therefore coupled into the governing equation. The variational multiscale method offers the viability of modeling the high Reynolds number flow. Hence, the intrusive Polynomial Chaos Expansion method is build upon this modelization for quantifying the impact of different uncertain parameters simultaneously.

Besides mathematical modeling, we place also our interest on the solver/preconditioner technique as solving such a complex system is challenging in practice. We therefore employ the inexact Multilevel preconditioner based on the comparison in our last paper \cite{Schick2015a} in a large system, we benefit from the low accuracy smoothing process to save computational effort.

Our following work will be focused on developing more analysis about the device's performance respect to surgical requirements. Further process of post-processing of our numerical simulation data and drawing more insight information about the uncertainties.

\section{ACKNOWLEDGMENT}
\label{sec:acknoledgement}

The results of this work is computed on bwForCluster MLS\&WISO (production and development). The authors acknowledge support by the state of Baden-Württemberg through bwHPC
and the German Research Foundation (DFG) through grant INST 35/1134-1 FUGG. The authors acknowledge support by the Heidelberg Institute for Theoretical Studies, HITS gGmbH.
 
\bibliographystyle{plain}
\bibliography{reference}

\end{document}